\documentclass[twocolumn]{aastex631}

\usepackage{xspace}
\newcommand{\Funit}{\ensuremath{\rm erg~s^{-1}~cm^{-2}}\xspace}

\newcommand{\Msun}{M_\odot}

\newcommand{\tcb}{}

\usepackage{graphicx,grffile,ulem,url,threeparttable,xcolor,soul,booktabs,natbib,amsmath,mathrsfs}
\usepackage{txfonts,rotating,longtable, needspace,subfigure,multirow,array}

\def\ne{\ensuremath{n_{\rm e}}\xspace}
\def\te{\ensuremath{T_{\rm e}}\xspace}
\newcommand\MSAEXP{\textsc{MSAEXP}\xspace}
\newcommand{\Ha}{\textrm{H}\ensuremath{\alpha}\xspace}
\newcommand{\Hb}{\textrm{H}\ensuremath{\beta}\xspace}
\newcommand{\Hg}{\textrm{H}\ensuremath{\gamma}\xspace}

\newcommand{\HII}{\textrm{H}\,\textsc{ii}\xspace}

\newcommand{\OII}{[\textrm{O}~\textsc{ii}]\xspace}
\newcommand{\OIII}{[\textrm{O}~\textsc{iii}]\xspace}
\newcommand{\CIII}{[\textrm{C}~\textsc{iii}]\xspace}

\newcommand{\SII}{[\textrm{S}~\textsc{ii}]\xspace}

\newcommand{\Ntot}{34\xspace}
\newcommand{\Novl}{13\xspace}
\newcommand{\Noii}{12\xspace}
\newcommand{\Nsii}{8\xspace}

\newcommand{\mstar}{$M_{*}$\xspace}

\shorttitle{NIRSpec glass - electron density}
\shortauthors{Li et al.}

\begin{document}

\title{Early Results from GLASS-JWST. XXV. Electron Density in the Interstellar Medium at $0.7\lesssim z\lesssim 9.3$ with NIRSpec High-resolution Spectroscopy\footnote{Based on observations acquired by the JWST under the ERS program ID 1324 (PI T. Treu)}}

\correspondingauthor{Xin Wang, Yuguang Chen}
\email{xwang@ucas.ac.cn, yugchen@ucdavis.edu}


\author[0000-0003-4813-8482]{Sijia Li}
\affil{School of Astronomy and Space Science, University of Chinese Academy of Sciences (UCAS), Beijing 100049, China}
\affil{Department of Astronomy, Xiamen University, Xiamen, Fujian 361005, China}

\author[0000-0002-9373-3865]{Xin Wang}
\affil{School of Astronomy and Space Science, University of Chinese Academy of Sciences (UCAS), Beijing 100049, China}
\affil{National Astronomical Observatories, Chinese Academy of Sciences, Beijing 100101, China}
\affil{Institute for Frontiers in Astronomy and Astrophysics, Beijing Normal University,  Beijing 102206, China}

\author[0000-0003-4520-5395]{Yuguang Chen}
\affiliation{Department of Physics and Astronomy, University of California Davis, 1 Shields Avenue, Davis, CA 95616, USA}

\author[0000-0001-5860-3419]{Tucker Jones}
\affiliation{Department of Physics and Astronomy, University of California Davis, 1 Shields Avenue, Davis, CA 95616, USA}

\author[0000-0002-8460-0390]{Tommaso Treu}
\affiliation{Department of Physics and Astronomy, University of California, Los Angeles, 430 Portola Plaza, Los Angeles, CA 90095, USA}

\author[0000-0002-3254-9044]{Karl Glazebrook}
\affiliation{Centre for Astrophysics and Supercomputing, Swinburne University of Technology, PO Box 218, Hawthorn, VIC 3122, Australia}


\author[0000-0002-1336-5100]{Xianlong He}
\affil{School of Astronomy and Space Science, University of Chinese Academy of Sciences (UCAS), Beijing 100049, China}
\affiliation{School of Physics and Technology, Wuhan University (WHU), Wuhan 430072, China}

\author[0000-0002-6586-4446]{Alaina Henry}
\affiliation{Space Telescope Science Institute, 3700 San Martin Drive, Baltimore MD, 21218} 
\affiliation{Center for Astrophysical Sciences, Department of Physics and Astronomy, Johns Hopkins University, Baltimore, MD, 21218}

\author[0009-0006-0596-9445]{Xiao-Lei Meng}
\affil{National Astronomical Observatories, Chinese Academy of Sciences, Beijing 100101, China}

\author[0000-0002-8512-1404]{Takahiro Morishita}
\affiliation{IPAC, California Institute of Technology, MC 314-6, 1200 E. California Boulevard, Pasadena, CA 91125, USA}

\author[0000-0002-4140-1367]{Guido Roberts-Borsani}
\affiliation{Department of Physics and Astronomy, University of California, Los Angeles, 430 Portola Plaza, Los Angeles, CA 90095, USA}

\author[0000-0002-8434-880X]{Lilan Yang}
\affiliation{Kavli Institute for the Physics and Mathematics of the Universe, The University of Tokyo, Kashiwa, Japan 277-8583}

\author[0000-0001-5277-4882]{Hao-Ran Yu}
\affil{Department of Astronomy, Xiamen University, Xiamen, Fujian 361005, China}


\author[0000-0003-2536-1614]{Antonello Calabr\`o}
\affiliation{INAF Osservatorio Astronomico di Roma, Via Frascati 33, 00078 Monteporzio Catone, Rome, Italy}

\author[0000-0001-9875-8263]{Marco Castellano}
\affiliation{INAF - Osservatorio Astronomico di Roma, via di Frascati 33, 00078 Monte Porzio Catone, Italy}

\author[0000-0003-4570-3159]{Nicha Leethochawalit}
\affiliation{National Astronomical Research Institute of Thailand (NARIT), MaeRim, Chiang Mai, 50180, Thailand}

\author[0000-0002-8632-6049]{Benjamin Metha}
\affiliation{School of Physics, The University of Melbourne, VIC 3010, Australia}

\author[0000-0003-2804-0648 ]{Themiya Nanayakkara}
\affiliation{Centre for Astrophysics and Supercomputing, Swinburne University of Technology, PO Box 218, Hawthorn, VIC 3122, Australia}

\author[0000-0002-4430-8846]{Namrata Roy}
\affiliation{Center for Astrophysical Sciences, Department of Physics and Astronomy, Johns Hopkins University, Baltimore, MD, 21218}

\author[0000-0003-0980-1499]{Benedetta Vulcani}
\affiliation{INAF Osservatorio Astronomico di Padova, vicolo dell'Osservatorio 5, 35122 Padova, Italy}

\newcommand{\approptoinn}[2]{\mathrel{\vcenter{
  \offinterlineskip\halign{\hfil$##$\cr
    #1\propto\cr\noalign{\kern2pt}#1\sim\cr\noalign{\kern-2pt}}}}}

\newcommand{\appropto}{\mathpalette\approptoinn\relax}

\begin{abstract}

The electron density (\ne) of the interstellar medium (ISM) in star-forming galaxies is intimately linked to star formation and ionization condition. Using the high-resolution spectra obtained from the JWST NIRSpec micro shutter assembly (MSA) as part of the GLASS-JWST program, we have assembled the largest sample to date (\Ntot galaxies) with individual \ne\ measurements derived from the \OII~$\lambda\lambda$3726,29 and/or \SII~$\lambda\lambda$6718,32 doublets at $0.7\lesssim z\lesssim 9.3$. The gravitational lensing magnification by the foreground Abell~2744 cluster allows us to probe {\ne in galaxies} with stellar masses (\mstar) down to $\simeq 10^{7.5} \Msun$ across the entire redshift range. Our analysis reveals that the \OII\ flux ratios are marginally anti-correlated with specific star formation rate (sSFR) within a 1-$\sigma$ confidence interval, whereas the \SII\ flux ratios show no significant correlation with sSFR. Despite clear correlation between sSFR and redshift within our sample, we find no apparent redshift evolution of \ne\ at $z \simeq 1$--9. Our dataset also includes \Novl\ galaxies where \ne\ can be measured from both \OII\ and \SII. Contrary to findings at lower redshifts, we observe considerable scatter in \ne\ measurements from \OII\ and \SII, indicating a complex gaseous environment with significant variations in \ne\ in high-redshift galaxies. This work highlights the unique capability of JWST NIRSpec/MSA high-resolution spectroscopy to characterize the detailed physical properties of the ISM in individual high-redshift galaxies. 

\end{abstract}

\section{Introduction}  \label{sec:intro}

Constraining the detailed properties of the interstellar medium (ISM) is paramount for a comprehensive understanding of the underlying physical mechanisms that drive the evolution of galaxies. As pristine gas from the intergalactic medium is accreted into galaxies, it cools and subsequently fuels star formation. However, intense star formation instigates gas outflows, which, through feedback mechanisms, enrich the ISM and regulate both the cooling and gas accretion processes \citep{lillyGASREGULATIONGALAXIES2013,2014_hopkins,Kewley2019,Harikane2020,Berg2021}. These intertwined processes, modulated by the ISM, concurrently influence the structure, dynamics, and chemical composition of the ISM through multifaceted physical activities such as heating and cooling, radiactive transfer, and turbulence \citep[see the review by][]{Klessen.2016}. Given this intricate relationship, it is of great interest to probe gas properties, such as temperature, density, and ionization states, and their variation over cosmic times from direct observational data \citep{Nagao2012,Nakajima2014,Feltre2016,Steidel2016,Sanders2016,Schaerer2018}. The density of free electrons ($n_e$) is particularly crucial, because it potentially reflects the gas density of ionized \HII regions and the evolution of photoionization, which are directly linked to the galaxy evolution \citep{isobe2023redshift,reddyJWSTNIRSpecExploration2023,2017FerlandRMxAA..53..385F}.   In addition, $n_e$ influences the brightness of emission lines with low critical densities, e.g., \OIII $\lambda$88~$\mu$m, that are often used to study galaxies at $z > 6$ \citep{Fujimoto2024, 2024_Schouws}. 

In H\,\textsc{ii} regions, $n_e$ can be determined by analyzing the intensity ratios of collisionally excited emission lines from specific ionized atoms. These lines originate from upper energy levels with similar excitation energy but different critical densities \citep{2006Osterbrock}. Commonly used intensity ratios include \OII$\lambda\lambda$3726/3729 \citep{2011_Rigby,isobe2023redshift,2023Reddy_Sigmasfr&U_MOSFIRE,Isobe2022}, \SII$\lambda\lambda$6716/6731 \citep{2008_Brinchmann_mn,2008_Liu_Xin,2009_Hainline,2010_Bian_Fuyan,2023_Miranda_Abundances}, \CIII$\lambda\lambda$1907/1909 \citep{2004_Rubin,2009_Quider,Kewley2019,Fujimoto2024}, and \OIII52$\mu $m/88$\mu $m \citep{Killi_2023, chen23}. Given the variety of \ne\ tracers, it is essential to understand whether these lines ratios yield consistent values. Recent study by \cite{2022Mingozzi} found that UV line ratios typically result in \ne\ values that are 1--2\,dex higher than those obtained from optical lines, \tcb{as the UV lines have different critical densities and probe different layers of the H\,\textsc{ii} regions.}
\tcb{\SII and \OII have by far the most extensive density measurements from normal $z \sim 0$ star forming galaxies. Adding [SII] and [OII] densities at high-$z$, as studied in this work, is therefore critical because those are the density diagnostics that have a good $z\sim 0$ benchmark. }

Spectroscopic surveys have extended measurement of \ne\ in ISM, using emission line flux ratios, to redshifts of approximately $z\sim9$ \citep{2008_Brinchmann_mn,2008_Liu_Xin,2009_Hainline,2009_Quider,2010_Bian_Fuyan,2011_Rigby,2014_Bayliss,Kewley2019,Fujimoto2024,isobe2023redshift,reddyJWSTNIRSpecExploration2023,stromCECILIAFaintEmission2023a}.
It is widely accepted that the \ne\ of the ISM in normal star-forming galaxies are generally higher in high-$z$ galaxies compared their local counterparts, sometimes by a factor of up to 100 \citep{2004_Rubin,2008_Brinchmann_mn,2008_Liu_Xin,2009_Hainline,2009_Quider,2010_Bian_Fuyan,2014_Shirazi_StarBornDenser,2014Shirazi_SINFONI,Kewley2019,2023Reddy_Sigmasfr&U_MOSFIRE,2023_Miranda_Abundances}. 
Leveraging the superior sensitivities and resolutions of JWST and Atacama Large Millimeter/submillimeter Array (ALMA),  \citet{Fujimoto2024} derived $n_e({\rm \OIII}) = 220^{+170}_{-100} \ \rm cm^{-3}$ for a lensed galaxy at $z = 8.496$ using the flux ratio of \OIII $\lambda$5007 and \OIII $\lambda$88$\mu m$. 
By analyzing medium resolution ($R = \lambda /{\Delta \lambda} \sim 1000$) JWST Near-Infrared Spectrograph (NIRSpec) spectra that marginally resolve the \OII\ $\lambda\lambda3726,29$ doublet, \cite{isobe2023redshift} reported $n_e \geq 300 \ \rm cm^{-3}$ in the ISM of the $z=4$--$9$ star-forming galaxies and found that \ne increases with higher redshift at a given $M_{*}$, star-formation rate (SFR), and specific SFR (sSFR). They also concluded that the variation of $n_e$ as a function of redshift is consistent with the size evolution of star-forming galaxies. This redshift evolution of $n_e$ may be attributed to the more compact H\,\textsc{ii} regions in high-redshift galaxies.  Alternatively, \cite{Kaasinen2017} suggested that the higher $n_e$ in many high-$z$ galaxies are closely linked to their elevated SFRs. From this perspective, the observed large $n_{\rm e}$ in high-$z$ galaxies may stem from a selection effect, as detections are biased towards galaxies undergoing intense star formation. Furthermore, \cite{2020_Harshan} found that protocluster galaxies at $z \sim 2$ may have higher \ne\ compared to the field galaxies, suggesting that denser galaxy environments enhance the ISM density.

The GLASS-JWST Early Release Science (ERS) program \citep[JWST-ERS-1324; PI: Treu;][]{Treu2022} provides an excellent opportunity that extends {the study of \ne} to low $M_{*}$ because of the lensing magnification introduced by the foreground galaxy cluster Abell 2744 (A2744). The total exposure time of NIRSpec is 17.7 kiloseconds with a spectral coverage of $1$ -- $5 \mu m$ from three \emph{high-resolution} $R \sim 2700$ gratings. In this work, we compile a sample of \Ntot star-forming galaxies at redshifts up to 9.3, with a lower limiting $M_*$ of 10$^{7.5}$ $M_{\odot}$. This sample doubles the amount of $z>1$ galaxies with individually measured \ne, allowing us to explore the relationship between \ne and host-galaxy properties (e.g., $z$, $M_*$, and SFR) with high fidelity.

This paper is organized as follows. In Section~\ref{sec:datsam} we describe the observations, sample selection, and data processing. We present the measurements of electron densities and other galaxy properties in Section~\ref{sec:analysis}. The discussion is provided in Section~\ref{sec:discu} and the conclusions are summarized in Section~\ref{sec:conclu}. Throughtout this work, we use AB magnitudes and assume $H_0=70~\rm \ km \ s^{-1}\ Mpc^{-1}$, $\Omega_M=0.3$, and $\Omega_\Lambda = 0.7$.

\begin{figure*}
    \centering
    \includegraphics[width=0.8\linewidth]{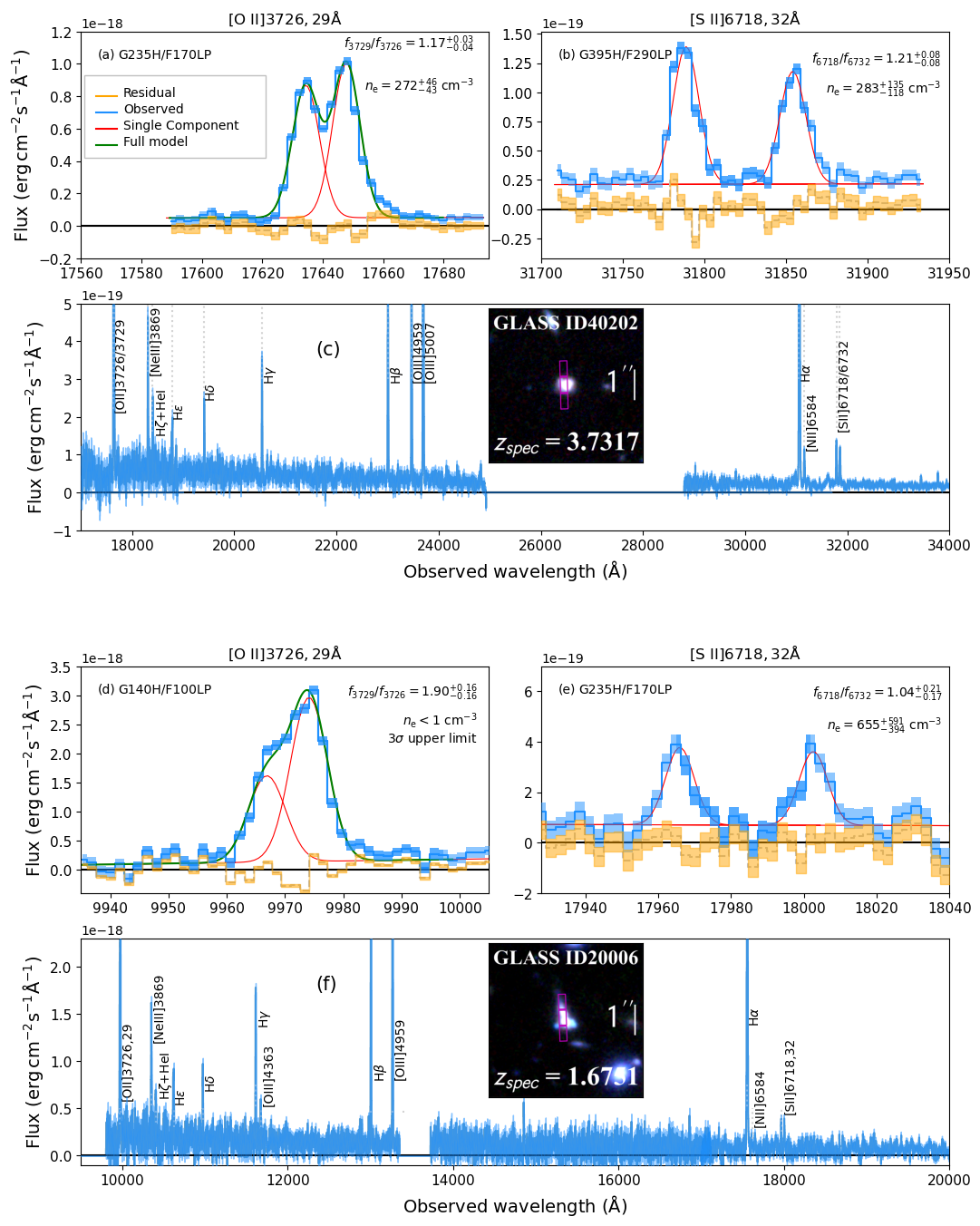}
    \caption{ Observed spectra and emission line fitting for GLASS-40202 at $z = 3.7317$ (top) and GLASS-20006 at $z = 1.6751$ (bottom). In both cases, the top two panels illustrate the emission line fits for the \OII and \SII doublets. The blue stepped lines represent the observed spectra, with the {blue boxes} indicating 1-$\sigma$ uncertainties. The best-fit single Gaussian for each line is shown in red, while the green curves represent the best-fit double Gaussian model for the doublets. The orange steps display the residuals of the fit. The bottom panels present the full observed spectrum of each galaxy, with prominent emission lines labeled. Inset images show color-composite images of the galaxies.
    }
    \label{fig.40202}
\end{figure*}

\section{Spectroscopic data and galaxy sample} \label{sec:datsam}

The multi-object spectroscopic data analyzed in this work are acquired from the GLASS-JWST program \citep{Treu2022} under the NIRSpec micro-shutter assembly (MSA) mode. The exposures are taken using high-resolution gratings of G140H/F100LP, G235H/F170LP, and G395H/F290LP, with an approximate resolving power of $R \approx 2700$. For each grating, an exposure time of 5 hours is used. The NIRSpec instrument has high sensitivity, such that emission lines \Ha, \Hb, \OII doublets, and {\OIII $\lambda5007$} can be detected at $> 3\sigma$ for a bright galaxy with $m_{\rm F150W} \lesssim 26.5\,$mag. The NIRSpec/MSA spectra cover a wavelength range of $\sim 0.6$--$5.3~\mu\mathrm{m}$, corresponding to $z \approx 1.68$--$12.40$ for the \OII doublets, and $z\approx 0.49$--$6.42$ for the \SII doublets.
The spectroscopic redshifts ($z_{\rm spec}$) of our sample galaxies are determined based on the strong nebular emission lines \OIII $\lambda5007$ after subtracting the stellar continuum.

For the data reduction of the NIRSpec/MSA spectra, the standard STScI JWST pipeline and the {\tt MSAEXP}\footnote{https://github.com/gbrammer/msaexp} software are employed, largely following the steps described in \citet{Jones_2023}. Initially, we generate the count-rate maps from the uncalibrated data using the \textsc{CALWEBB\_DETECTOR1} module, incorporating up-to-date reference files (jwst\_1040.PMAP) for accurate calibration. Subsequently, the {\tt MSAEXP} software performs additional preprocessing steps to eliminate the 1/f noise and ``snowball" features observed in the rate images. To extract the two-dimensional (2D) spectra from individual exposures, we employ the level-2 \textsc{CALWEBB\_SPEC2} reduction scripts, which involved World Coordinate System (WCS) registration, slit path-loss correction, flat-fielding, wavelength calibration, and flux calibration. These steps ensure the precise alignment, correction, and calibration of the extracted 2D spectra. Then, leveraging the \citet{Horne_1986} algorithm, {\tt MSAEXP} performs an optimal one-dimensional (1D) spectral extraction. This algorithm uses the target light profile along the direction of cross-dispersion to determine the optimal extraction aperture, resulting in an enhanced signal-to-noise ratio in the final 1D spectra. Overall, our data reduction approach that combines the STScI JWST pipeline and \MSAEXP enables accurate calibration, extraction, and optimization of the NIRSpec spectra, facilitating detailed analysis and scientific investigations of celestial objects. Finally, to enhance the overall quality and reliability of the data, the 1D spectra obtained from multiple exposures taken at different dither positions and visits are merged using median stacking to remove any outliers.

From the extensive GLASS-JWST NIRSpec dataset, we identified \Ntot sources with \SII or \OII doublets significantly detected with signal-to-noise ratios (SNRs) greater than 3. The details of our defined sample can be found in Table~\ref{table:1}. Among these \Ntot sources, \Noii of them have only \OII covered (the ``\OII'' sample), \Nsii of them have only \SII covered (the ``\SII'' sample), and the rest \Novl galaxies have both \SII and \OII doublets detected at high significance with SNR $>$3 (the ``overlap'' sample).
The ``overlap'' sample is, to our knowledge, the first sample at high redshifts $( 1.6 \lesssim z \lesssim 3.7 )$ where the ISM electron density can be probed using both tracers.

Figure~\ref{fig.40202} showcases two example galaxies in the overlap sample, with both \OII $\lambda\lambda$3726,29, and \SII$ \lambda\lambda$6718,32 doublets detected with $\mathrm{SNRs} \gtrsim 20$. In previous works, \ne\ measured in similar redshift ranges often relied on stacking techniques to achieve a comparable SNR. 
The redshift of our entire galaxy samples ranges from 0.7 to 9.3, while their $M_*$ covers $\log(M_*/M_{\odot})\approx 7.5$--10.4 derived from the spectral energy distribution (SED) fitting {(see Section~\ref{subsec:m_sfr} for details)}. The samples span two orders of magnitude in SFR from 0.36 to  23.25$\,M_{\odot}\,{\rm yr}^{-1}$.

\section{Analysis and Measurements}  \label{sec:analysis}

\subsection{Line Flux}   \label{subsec:fit_ne}

Physical properties \ne, electron temperature (\te), and SFR are derived from emission line fluxes or flux ratios. We use {\tt LIME}\footnote{\url{https://github.com/Vital-Fernandez/lime}} \citep{2024LIME} to measure the emission line fluxes and associated errors. {\tt LIME} fits emission lines using Gasussian functions through non-linear least-squares minimization, which is implemented by \textsc{lmfit}\footnote{\url{https://lmfit.github.io/lmfit-py/}}.  

For each emission line, {\tt LIME} calculates the continuum surrounding the line and sets the fitting wavelength range to $\pm 50\,\mathrm{\AA}$ around the line in the rest wavelengths. However, for a few \Ha\ lines, which exhibit faint broad components possibly due to stellar winds, the fitting region is further constrained to the central Gaussian components. This adjustment effectively reduces contamination. We note that the \Ha lines are significantly stronger than these broad components, indicating the impact of the broad components in the narrow central region is minimal and does not significantly adversely affect our results. {This faint broad component could be due to the Raman scattering effect of \Ha as discovered in observations of \HII regions in the LMC, SMC and Orion (M42) \citep{2016_Dopita}.}

We use {\tt LIME} to fit a Gaussian function to \OIII$\lambda 5007$, which is relatively strong across all our galaxies, to determine redshift. Subsequently, we fit Gaussian functions to \Ha, \Hb, \Hg, and \OIII$\lambda 5007$ lines. For the \OII and \SII doublets, we apply a double Gaussian fit, where the two functions are fitted simultaneously, sharing the same velocity dispersion ($\sigma$). The flux of a line is defined as the integrated flux of the best-fit Gaussian function, which the fitting error provided by {\tt LIME} taken as the uncertainty of the flux. The centroids of the lines are allowed to be adjust slightly during the fitting process, while the wavelength separation of the doublets is fixed to theoretical values. Figure~\ref{fig.40202} illustrates the fitting process. 

To accurately determine the SFR from Balmer lines {in Sect.}~\ref{subsec:balmer_sfr}, the total flux from the entire galaxy must be measured. However, slit spectroscopy inherently misses part of the galaxy light, making slit loss correction necessary to recover the total emission flux. We apply the default settings of the \textsc{pathloss} correction step in the \textsc{CALWEBB\_SPEC2} pipeline to account for slit losses during data reduction. To further correct for any residual slit loss and potential inaccuracies in flux calibration, we convert the extracted 1D spectra into synthetic fluxes as observed by JWST NIRCam, using the filter throughput. The slit loss correction factor is then calculated as the ratio of the total galaxy flux from the photometric catalog constructed by the Dawn JWST Archive (DJA)\footnote{https://dawn-cph.github.io/dja/index.html} to the synthetic flux.  The filter used for this calculation covers the Balmer line used for SFR determination. The correction factor is then multiplied by the measured Balmer line fluxes to obtain the slitloss-corrected fluxes. The DJA galaxy fluxes are also used for SED fitting to obtain stellar mass in Sect.~\ref{subsec:m_sfr}. No slit loss correction is applied to other lines, such as the \OII\ and \SII\ doublets, as their flux ratios, rather than absolute fluxes, are used for deriving electron properties. Table~\ref{table:1} summarizes the measured redshifts, fluxes, and associated uncertainties.

{
\tabletypesize{\scriptsize}
\begin{deluxetable*}{llllcccccccccc}   \tablecolumns{10}
\tablewidth{0pt}
\tablecaption{Observed properties of individual sources in our galaxy samples with ISM electron density (\ne) measurements.}
\tablehead{
    \colhead{ID} & 
    \colhead{R.A.} & 
    \colhead{Dec.} & 
    \colhead{$z_\mathrm{spec}$} &  
    \colhead{$f_{\rm [OII] \lambda3726}$} & 
    \colhead{$f_{\rm [OII] \lambda3729}$} & 
    \colhead{$f_{\rm [SII] \lambda 6718}$} & 
    \colhead{$f_{\rm [SII] \lambda 6732}$} &
    \colhead{$f_{\rm H\alpha }$} &
    \colhead{$f_{\rm H\beta }$} & 
    \\& \multicolumn{2}{c}{\hrulefill} &  &\multicolumn{6}{c}{\hrulefill}\\
    &\multicolumn{2}{c}{[deg]} &  & \multicolumn{6}{c}{{Observed emission line fluxes [$10^{-19}$\Funit]}} 
}
\startdata
    \multicolumn{10}{c}{overlap sample}   \\
    \noalign{\smallskip} 
GLASS-20006 & 3.571763	& $-30.380432$	&1.675	 &$119\pm   9 $&$   225 \pm  8 $ &  $30  \pm  4 $&$  30  \pm  4 $&$ 698 \pm 9    $&$ 234 \pm 4     $&\\                       
GLASS-20008&  3.570194	& $-30.383723$	&1.8631	 &$167 \pm  4 $& $ 195 \pm  3  $ &  $52  \pm  2 $&$ 43   \pm 2  $&$ 618 \pm 10   $&$ 144 \pm 3     $&\\
GLASS-20025&  3.586964	& $-30.386998$  &1.8626	 &$ 39 \pm  3 $& $  65 \pm  3  $ &  $11  \pm  1 $&$  10  \pm  1 $&$ 113 \pm 2    $&$ 39  \pm 2     $&\\
GLASS-40094	& 3.604914  & $-30.418699$  &2.6745  &$19 \pm   7 $&$   53 \pm  7  $ &  $15  \pm  4 $&$  21  \pm  4 $&$ 169 \pm 4    $&$    $&\\
GLASS-40202&  3.616540	& $-30.424521$	&3.7317	 &$ 94 \pm  2 $& $ 110 \pm  2  $ &  $24  \pm  1 $&$  20  \pm  1 $&$ 355 \pm 4    $&$ 95  \pm 1     $&\\
GLASS-80027&  3.569293   & $-30.409627$	&3.5797	 &$ 14 \pm  2 $& $  20 \pm  2  $ &  $ 3.4  \pm  0.8 $&$  3.2   \pm  0.8 $&$ 142 \pm 3    $&$    $&\\
GLASS-320106& 3.581208	& $-30.421063$  &2.5083	 &$ 57 \pm  2 $& $  64 \pm  1  $ &  $20  \pm  1 $&$  13  \pm  1 $&$ 246 \pm 3    $&$ 89  \pm 4     $&\\
GLASS-320027& 3.573612  & $-30.427573$	&1.6889	 &$ 35 \pm  7 $& $  36 \pm  4  $ & $321  \pm  14 $&$ 298 \pm  14$&$ 1690 \pm 120 $&$    $&\\
GLASS-340899& 3.566306	& $-30.378432$	&2.3437	 &$119 \pm  4 $& $ 155 \pm  4  $ &  $22  \pm  2 $&$  21  \pm  2 $&$ 443 \pm 5    $&$ 164  \pm 7     $&\\
GLASS-340920& 3.605020  & $-30.375960$  &2.4874  &$12 \pm   1 $&$   14 \pm  1 $  &  $8.3	\pm  1.7 $&$  5.7   \pm  1.6 $&$ 45 \pm 2    $&$    $&\\
GLASS-341691& 3.561206	& $-30.419820$	&2.3717	 &$ 34 \pm  1 $& $  48 \pm  1  $ &  $ 4.7  \pm  0.6 $&$  2.9   \pm  0.5 $&$ 138 \pm 1    $&$ 58  \pm 2     $&\\
GLASS-342363& 3.581425	& $-30.429338$	&1.7567	 &$221 \pm  21 $& $ 254 \pm 17 $ & $122  \pm 12 $&$ 107  \pm  12$&$ 1070 \pm 20    $&$ 145 \pm 10     $&\\
GLASS-342371& 3.619580	& $-30.429116$	&2.5820	 &$117 \pm  3 $& $ 142 \pm  3  $ &  $27  \pm  2 $&$  24  \pm  2 $&$ 1020 \pm 10    $&$ 318 \pm 5     $&\\
GLASS-410044& 3.612807	& $-30.390217$  &1.7285	 &$ 26 \pm  2 $& $  42 \pm  2  $ &  $ 8.7  \pm  1.8 $&$  6.5   \pm  1.5 $&$ 121 \pm 2    $&$ 31  \pm 1     $&\\
\noalign{\smallskip}\hline\noalign{\smallskip}
\multicolumn{10}{c}{\OII sample}   \\
\noalign{\smallskip}
GLASS-10003&  3.617162	& $-30.425545$	&9.3127	 &$  13 \pm  2 $& $  14 \pm  2  $ &			           &                     &$  $&$    $&\\
GLASS-10021	& 3.608511  & $-30.418541$  &7.2863  &$  25 \pm   3 $&$   35 \pm  2  $ &                     &                     &$  $&$ 74  \pm 2     $&\\
GLASS-40050	& 3.579843  & $-30.426287$  &2.9845  &$  58 \pm   6 $&$   46 \pm  6  $ &                     &                     &$ 109 \pm 4    $&$ 29  \pm 4     $&\\
GLASS-50038&  3.565199	& $-30.394264$	&5.7720	 &$  8.3 \pm  1.1 $& $  14 \pm  1  $ &			           &                     &$ 88 \pm 10    $&$ 17  \pm 2     $&\\
GLASS-100001& 3.603845  & $-30.382234$	&7.8732	 &$  8.9 \pm  1.2 $& $  13 \pm  1  $ &			           &                     &$  $&$ 8    \pm 1     $&\\
GLASS-110000& 3.570642	& $-30.414638$	&5.7641	 &$  9.3 \pm  1.0 $& $   6.0 \pm  0.8  $ &			           &                     &$ 21 \pm 1    $&$ 10  \pm 1     $&\\
GLASS-150029& 3.577166  & $-30.422576$	&4.5838	 &$  12 \pm  1 $& $  17 \pm  1  $ &			           &                     &$ 186 \pm 3    $&$ 50  \pm 1     $&\\
GLASS-160122& 3.564901	& $-30.424956$	&5.3319	 &$  4.4 \pm  1.1 $& $   4.8 \pm  1.1  $ &			           &                     &$ 40 \pm 1    $&$ 18  \pm 1     $&\\
GLASS-160248& 3.613190	& $-30.411679$	&3.0520	 &$  33 \pm  2 $& $  40 \pm  2  $ &			           &                     &$  $&$ 48  \pm 3     $&\\
GLASS-340935& 3.605068  & $-30.376649$  &1.6506  &$  39 \pm   5 $&$   47 \pm  5  $ &                     &                     &$ 13.3 \pm 0.5    $&$ 32  \pm 3     $&\\
GLASS-342321& 3.619052	& $-30.431628$	&2.7100	 &$  83 \pm  5 $& $ 218 \pm  5  $ &			           &                     &$ 589 \pm 7    $&$ 208 \pm 4     $&\\
GLASS-360007& 3.577464	& $-30.410947$	&3.2055	 &$  10 \pm  1 $& $  15 \pm  1  $ &			           &                     &$ 44 \pm 1    $&$ 17  \pm 1     $&\\
\noalign{\smallskip}\hline\noalign{\smallskip}
\multicolumn{10}{c}{\SII sample}   \\
\noalign{\smallskip} 
GLASS-20021 & 3.576740	& $-30.393605$	&1.3667	 &     				 &                       &  $63  \pm  3 $&$  47  \pm  3 $&$ 834 \pm 13    $&$ 250 \pm 4     $&\\
GLASS-40054 & 3.569485	& $-30.427003$	&2.5796	 &     				 &                       &  $ 2.5  \pm  0.4 $&$  1.9   \pm  0.4 $&$ 20 \pm 1    $&$ 7   \pm 1     $&\\
GLASS-40066 & 3.599716	& $-30.431894$	&4.0199	 &     				 &                       &  $21  \pm  2 $&$  10  \pm  1 $&$ 425 \pm 5    $&$ 143  \pm 2     $&\\
GLASS-80029 & 3.603180  & $-30.415709$	&3.9513	 &     				 &                       &  $ 6.0  \pm  1.2 $&$  5.5   \pm  1.2 $&$ 117\pm 1    $&$ 31  \pm 1     $&\\
GLASS-160133& 3.580275	& $-30.424404$	&4.0165  &                   &                       &  $5.0  \pm	 0.9 $& $3.3    \pm 0.9 $ &$ 526 \pm 11    $&$ 8   \pm 1     $&\\
GLASS-341721& 3.562174	& $-30.416355$	&2.5605	 &     				 &                       &  $20  \pm  1 $&$  16  \pm  1 $&$ 275 \pm 5    $&$    $&\\
GLASS-410024& 3.579066	& $-30.395852$	&0.6879	 &     				 &                       &  $44  \pm  2 $&$  37  \pm  2 $&$ 396 \pm 5    $&$    $&\\
GLASS-410063& 3.608077  & $-30.377501$  &0.9614  &		             &                       &  $67 \pm  4 $&$  42  \pm  4 $&$ 536 \pm 9    $&$    $&\\
GLASS-410067& 3.586249	& $-30.416541$	&1.2694	 &     				 &                       &  $ 7  \pm  1 $&$  4.2   \pm  0.9 $&$ 66 \pm 1    $&$ 16  \pm 1     $&\\
\enddata

\tablecomments{
    R.A. and Dec. are the right ascension and declination in J2000, respectively. 
    The line fluxes presented in this table are measured from the observed spectra without the slitloss correction, reddening correction, or lensing magnification correction.
    }
\label{tab:flux}
\end{deluxetable*}
}  \label{table:1}
{
\tabletypesize{\scriptsize}
\begin{deluxetable*}{lccccc}
\tablecolumns{6}
\tablewidth{0pt}
\tablecaption{Derived physical properties of individual sources in our galaxy samples with ISM electron density (\ne) measurements.}
\tablehead{
    \colhead{ID} & 
    \colhead{$n_e({\rm [SII] })$} &
    \colhead{$n_e({\rm [OII] })$} & 
    \colhead{$\log(M_{*}/M_{\odot})$} & 
    \colhead{$\rm Av_{SED}$} & 
    \colhead{$\rm SFR_{Balmer}$} \\
     & \multicolumn{2}{c}{\hrulefill} &  & &\multicolumn{1}{c}{\hrulefill} \\
    & \multicolumn{2}{c}{[cm$^{-3}$]}  & & & [$M_{\odot}$\,yr$^{-1}$] 
}
\startdata
    \multicolumn{6}{c}{overlap sample}   \\
    \noalign{\smallskip} 
GLASS-20006     & $660	_{-390}	^{+590}	$  &  $	\textless\,{0}     $             &$ 8.16 ^{+ 0.01 }_{- 0.01 }$&$ 0.15 ^{+ 0.01 }_{- 0.01 }$&$ 8.53 ^{+ 0.18 }_{- 0.18 }$ \\     
GLASS-20008     & $257	_{-83}	^{+93 } $  &  $267	    _{-33}	 ^{+38	 }$      &$ 9.06 ^{+ 0.0 }_{- 0.0 }$&$ 0.37 ^{+ 0.03 }_{- 0.03 }$&$ 13.0 ^{+ 0.5 }_{- 0.5 }$\\
GLASS-20025     & $520	_{-290}	^{+450}	$  &  $\textless\,          {0	 }$      &$ 7.85 ^{+ 0.01 }_{- 0.01 }$&$ 0.09 ^{+ 0.01 }_{- 0.01 }$&$ 1.99 ^{+ 0.04 }_{- 0.04 }$\\
GLASS-40094     & $2000_{-1300}^{+5600}$  &  $	\textless\,{0}     $             &$ 8.96 ^{+ 0.0 }_{- 0.0 }$&$ 0.0 ^{+ 0.0 }_{- 0.0 }$&$ 2.33 ^{+ 0.05 }_{- 0.05 }$\\
GLASS-40202     & $280	_{-120}	^{+140}	$  &  $272	    _{-43}	 ^{+46	 }$      &$ 8.97 ^{+ 0.02 }_{- 0.02 }$&$ 0.3 ^{+ 0.02 }_{- 0.03 }$&$ 22.5 ^{+ 0.6 }_{- 0.8 }$\\
GLASS-80027     & $590	_{-590}	^{+1500}$  &  $\textless\,        {148	 }$      &$ 8.44 ^{+ 0.01 }_{- 0.02 }$&$ 0.16 ^{+ 0.02 }_{- 0.02 }$&$ 2.45 ^{+ 0.08 }_{- 0.08 }$\\
GLASS-320106    & $\textless\,   {52}  $  &  $303	    _{-46}	 ^{+57	 }$       &$ 9.03 ^{+ 0.0 }_{- 0.0 }$&$ 0.2 ^{+ 0.02 }_{- 0.02 }$&$ 7.38 ^{+ 0.2 }_{- 0.2 }$\\
GLASS-320027    & $530	_{-150}	^{+160}	$  &  $470	    _{-300}	 ^{+450	 }$       &$ 10.52 ^{+ 0.0 }_{- 0.0 }$&$ 0.26 ^{+ 0.02 }_{- 0.02 }$&$ 21.1 ^{+ 1.9 }_{- 1.9 }$\\
GLASS-340899    & $620	_{-290}	^{+330}	$  &  $135	    _{-46}	 ^{+54	 }$       &$ 7.53 ^{+ 0.02 }_{- 0.02 }$&$ 0.39 ^{+ 0.03 }_{- 0.03 }$&$ 5.84 ^{+ 0.2 }_{- 0.19 }$\\
GLASS-340920    & $13	   _{-13}^{+590}$  & $280	    _{-190}	^{+230}$          &$ 7.92 ^{+ 0.01 }_{- 0.01 }$&$ 0.07 ^{+ 0.02 }_{- 0.03 }$&$ 0.73 ^{+ 0.04 }_{- 0.05 }$\\
GLASS-341691    & $\textless\,   {164}	$  &  $52	    _{-40}	 ^{+46	 }$       &$ 8.24 ^{+ 0.06 }_{- 0.06 }$&$ 0.2 ^{+ 0.04 }_{- 0.04 }$&$ 2.39 ^{+ 0.09 }_{- 0.09 }$\\
GLASS-342363    & $390	_{-260}	^{+390}	$  &  $300	    _{-150}	 ^{+170	 }$       &$ 8.95 ^{+ 0.01 }_{- 0.01 }$&$ 0.37 ^{+ 0.03 }_{- 0.03 }$&$ 3.08 ^{+ 0.13 }_{- 0.13 }$\\
GLASS-342371    & $400	_{-180}	^{+260}	$  &  $218	    _{-41}	 ^{+44	 }$       &$ 8.11 ^{+ 0.01 }_{- 0.01 }$&$ 0.18 ^{+ 0.01 }_{- 0.01 }$&$ 12.2 ^{+ 0.2 }_{- 0.2 }$\\
GLASS-410044    & $99	_{-99}	^{+670}	$  &  $\textless\,        {12	 }$       &$ 9.04 ^{+ 0.01 }_{- 0.01 }$&$ 0.03 ^{+ 0.02 }_{- 0.02 }$&$ 2.7 ^{+ 0.16 }_{- 0.16 }$\\
\noalign{\smallskip}\hline\noalign{\smallskip}
\multicolumn{6}{c}{\OII sample}   \\
\noalign{\smallskip}
GLASS-10003     & $                    $  &  $420	    _{-290}	 ^{+370	 }$  &$ 9.11 ^{+ 0.05 }_{- 0.05 }$&$ 0.03 ^{+ 0.02 }_{- 0.01 }$&$               $\\
GLASS-10021	    &                         & $68	_{-68}	^{+145}$             &$ 8.42 ^{+ 0.01 }_{- 0.01 }$&$ 0.12 ^{+ 0.01 }_{- 0.01 }$&$ 23.3 ^{+ 0.9 }_{- 0.9 }$\\
GLASS-40050	    &                         & $1090	_{-440}	^{+680}$         &$ 8.76 ^{+ 0.01 }_{- 0.01 }$&$ 0.1 ^{+ 0.01 }_{- 0.01 }$&$ 1.7 ^{+ 0.08 }_{- 0.08 }$\\
GLASS-50038     & $	                $  &  $\textless\,          {2	 }$      &$ 8.85 ^{+ 0.01 }_{- 0.01 }$&$ 0.13 ^{+ 0.02 }_{- 0.02 }$&$               $\\
GLASS-100001    & $	                $  &  $58	    _{-58}	 ^{+180	 }$       &$ 9.26 ^{+ 0.04 }_{- 0.07 }$&$ 0.42 ^{+ 0.05 }_{- 0.05 }$&$ 5.9 ^{+ 1.45 }_{- 1.32 }$\\
GLASS-110000    & $	                $  &  $1900     _{-730}	 ^{+1400 }$	      &$ 8.78 ^{+ 0.02 }_{- 0.02 }$&$ 0.17 ^{+ 0.02 }_{- 0.02 }$&$ 2.96 ^{+ 0.12 }_{- 0.12 }$\\
GLASS-150029    & $	                $  &  $65	    _{-65}	 ^{+88	 }$       &$ 8.31 ^{+ 0.02 }_{- 0.02 }$&$ 0.23 ^{+ 0.02 }_{- 0.02 }$&$ 6.92 ^{+ 0.26 }_{- 0.25 }$\\
GLASS-160122    & $	                $  &  $370	    _{-370}	 ^{+780  }$       &$ 8.41 ^{+ 0.16 }_{- 0.17 }$&$ 0.03 ^{+ 0.04 }_{- 0.02 }$&$ 1.62 ^{+ 0.08 }_{- 0.05 }$\\
GLASS-160248    & $	                $  &  $208	    _{-79}	 ^{+90	 }$       &$ 8.69 ^{+ 0.01 }_{- 0.01 }$&$ 0.25 ^{+ 0.02 }_{- 0.02 }$&$ 9.62 ^{+ 0.8 }_{- 0.79 }$\\
GLASS-340935    &                         & $260	_{-170}	^{+230}$          &$ 8.1 ^{+ 0.02 }_{- 0.03 }$&$ 0.0 ^{+ 0.01 }_{- 0.0 }$&$ 0.36 ^{+ 0.0 }_{- 0.0 }$\\
GLASS-342321    & $	                $  &  $\textless\,          {0	 }$       &$ 8.64 ^{+ 0.05 }_{- 0.04 }$&$ 0.27 ^{+ 0.05 }_{- 0.04 }$&$ 5.96 ^{+ 0.3 }_{- 0.25 }$\\
GLASS-360007    & $	                $  &  $32	    _{-32}	 ^{+140	 }$       &$ 9.02 ^{+ 0.0 }_{- 0.01 }$&$ 0.14 ^{+ 0.02 }_{- 0.02 }$&$ 2.47 ^{+ 0.11 }_{- 0.11 }$\\
\noalign{\smallskip}\hline\noalign{\smallskip}
\multicolumn{6}{c}{\SII sample}   \\
\noalign{\smallskip} 
GLASS-20021     & $100	_{-100}	^{+120}	$  &  	    	    	                 &$ 9.18 ^{+ 0.0 }_{- 0.0 }$&$ 0.0 ^{+ 0.0 }_{- 0.0 }$&$ 6.53 ^{+ 0.1 }_{- 0.1 }$\\
GLASS-40054     & $110	_{-110}	^{+460}	$  &  	    	    	                 &$ 9.24 ^{+ 0.05 }_{- 0.05 }$&$ 0.24 ^{+ 0.07 }_{- 0.07 }$&$ 3.4 ^{+ 0.28 }_{- 0.28 }$\\
GLASS-80029     & $440	_{-440}	^{+940}	$  &  	    	    	                 &$ 7.97 ^{+ 0.02 }_{- 0.02 }$&$ 0.21 ^{+ 0.02 }_{- 0.02 }$&$ 4.32 ^{+ 0.12 }_{- 0.12 }$\\
GLASS-160133    & $	\textless\,{489}  $&                                          &$ 7.83 ^{+ 0.0 }_{- 0.0 }$&$ 0.25 ^{+ 0.02 }_{- 0.02 }$&$ 10.7 ^{+ 0.4 }_{- 0.4 }$\\
GLASS-341721    & $190	_{-120}	^{+160}	$  &  	    	    	                  &$ 8.16 ^{+ 0.02 }_{- 0.02 }$&$ 0.04 ^{+ 0.02 }_{- 0.02 }$&$ 0.44 ^{+ 0.01 }_{- 0.01 }$\\
GLASS-410024    & $320	_{-150}	^{+160}	$  &  	    	    	                  &$ 8.67 ^{+ 0.01 }_{- 0.01 }$&$ 0.18 ^{+ 0.02 }_{- 0.01 }$&$ 0.74 ^{+ 0.02 }_{- 0.02 }$\\
GLASS-410063    & $\textless\,   {10}  $  &                                       &$ 9.45 ^{+ 0.0 }_{- 0.0 }$&$ 0.14 ^{+ 0.01 }_{- 0.01 }$&$ 2.8 ^{+ 0.06 }_{- 0.06 }$\\
GLASS-410067    & $\textless\,   {144}	$  &  	    	    	                  &$ 9.17 ^{+ 0.0 }_{- 0.0 }$&$ 0.0 ^{+ 0.0 }_{- 0.0 }$&$ 1.12 ^{+ 0.04 }_{- 0.04 }$\\
\enddata
    \tablecomments{
    \ne(\SII) and \ne(\OII) are the ISM electron density derived from the \SII or \OII doublets, respectively, assuming a uniform electron temperature of 10,000 K (see Section \ref{sec:analysis}).
    The error bars and upper limits reported in this table all correspond to 1-$\sigma$ confidence intervals.
    The measurements presented in this table have been corrected for lensing magnification, if applicable.
    }
\label{tab:ne}
\end{deluxetable*}  \label{table:2}

\subsection{Electron Density}   \label{subsec:fit_ne}
\textcolor{teal}{}

Assuming an electron temperature ($T_e$) of 10,000\,K, which is typical for galaxies at $z > 2$ \citep{Steidel2014, 2023Sanders, 2024Sanders}, we performed nebular analysis with the default atomic databases to derive the \ne\ using the \SII $\lambda\lambda6718/6732$ and/or \OII $\lambda\lambda3276/3729$ ratios via \textsc{PyNeb} \citep{Luridiana2015}. 

The relationship between the line ratio and \ne\ is nearly linear when \ne\ is near the critical density, but it flattens significantly when \ne\ deviates from the critical density \citep{DraineBook}. The theoretical upper limit is achieved when \ne\ is close to zero. 
Some \OII or \SII doublets exceed their theoretical upper limits, which are unphysical and cannot be explained by current models. A notable example is GLASS-20006, shown in Figure \ref{fig.40202}, where the ${\rm SNR_{\OII \lambda 3726}} = 13$ and ${\rm SNR_{\OII \lambda 3729}} = 29$, thanking to the high sensitivity of JWST. The best-fit models yield \OII $\lambda\lambda3729/3726 = 1.9 \pm 0.16$,  significantly above the theoretical upper limit of 1.4. The observed high ratio should not be attributed to flux calibration errors, as the doublets are in close proximity. 
{In these cases where the line ratio is 1-$\sigma$ above the theoretical upper limit, the method totally fails and \ne is denoted as $<0$.}
If the line ratio is higher than the theoretical upper limit, but the lower 1\,$\sigma$ value is below it, an upper limit for \ne is derived corresponding to the lower 1\,$\sigma$ value. 
\citet{Sanders2016} provided relations between \OII\ and \SII\ line ratios and \ne\  at 10,000\,K using updated atomic data. As a sanity check, we derived \ne\ from these relations and found them to be consistent with our values.

The calculation of \ne\ is largely insensitive to the input $T_e$ \citep{2006Osterbrock}. For one galaxy in our sample where \OIII$\lambda 4363$ was detected, we derive $T_e$ from the line flux (see next section) and used it to calculate \ne. The resulting \ne closely matched the value derived with $T_e=10,000\,K$, with a fractional difference of 5\%.

We note that, when \ne\ is significantly higher or lower than the critical values, the uncertainty in \ne propagated from the uncertainty of line ratio is a large value because of the flattening of the relation between \ne and line ratio. To estimate the uncertainty in \ne, we performed Monte Carlo (MC) simulations by randomly perturbing the \OII\ and \SII\ doublets fluxes based on their errors and rerunning \textsc{PyNeb} with the same setup. This process was repeated 10,000 times.  The lower and upper uncertainty in \ne\ were determined as the 16th and 84th percentiles of the recalculated \ne\ from the MC simulation. The derived \ne and their uncertainties are presented in Table~\ref{tab:ne}.

\subsection{Electron Temperature}   \label{subsec:Temeasurements}
Despite that the \OII\ and \SII\ doublets are almost only sensitive to \ne, thus, making them good \ne\ indicators, electron temperature (\te) can still have small effects on the \OII\ and \SII\ line ratios. To understand the systematic uncertainties associated with \te, we identified one galaxy (GLASS-40202) in our overlapping sample. This galaxy has the auroral line \OIII$\lambda 4363$ detected at $\mathrm{SNR} > 3$, and thus \te\ can be reliably measured. We find six galaxies in the overlapping sample with spectra available at the wavelength of the \OIII$\lambda 4363$ line and then calculate the upper flux limit of this line, which is used to estimate \te. Ideally, we would like to detect the \OII$\lambda7319,7330$ auroral lines, which provides the direct \te\ for the \OII\ gas. However, none of our galaxies in the overlapping sample have the \OII$\lambda7319,7330$ reaching the $\mathrm{SNR} > 3$ threshold. Therefore, we use the empirical $T_2$-$T_3$ relationship from \cite{PerezMontero17}, i.e., 

\begin{align}
    T_2 &= T_{e}[{\rm O}^{+}] = T_{e}[{\rm S}^{+}], \\
    T_3 &= T_{e}[{\rm O}^{++}], 
\end{align}
and, 
\begin{align}
    T_2 &= \frac{2\times10^4 K}{(0.8 + 10^4 K/T_3)}.
\end{align}

To estimate $T_{e}[{\rm O}^{++}]$, we adopt the empirical relation between $ R_{\rm O3} =(F_{4959}+F_{5007})/F_{4363}$, where $F_{4959}$, $F_{5007}$, and $F_{4363}$ are the flux of \OIII $\lambda$4959, \OIII $\lambda$5007, and \OIII $\lambda$4363, respectively, and \te from \cite{PerezMontero17}:
\begin{align}\label{eq.te}
    T_{e}[{\rm O}^{++}] & = 10^4 K * (0.7840-0.0001357 \times R_{\rm O3} +48.44/R_{\rm O3}) .
\end{align}
The $T_{e}[{\rm O}^{++}]$ is calculated from the dust-corrected $(F_{4959}+F_{5007})/F_{4363}$ ratio using the nebular $E(B-V)$ and the \cite{1994Calzetti} extinction curve in \textsc{PyNeb}. The resulting \te is then used to calculate \ne following Sect.~\ref{subsec:fit_ne} and used to discuss the impact of \te on \ne measurement in Sect.~\ref{subsec:oii_sii}.

\subsection{Unphysical Line Ratios and Their Detectability Using Medium Resolution Spectroscopy} \label{subsec:appendix}

To properly resolve the \OII$\lambda\lambda$3726,29 emission line doublets for accurate \ne measurements, a wavelength resolution of $\Delta\lambda\lesssim1.4~\AA$ (FWHM $\sim 120 km/s$) is required in the rest frame to separate the doublet peaks by at least two resolution elements. This is equivalent to $R\gtrsim2660$. The NIRSpec/MSA high-resolution spectroscopy obtained by GLASS-JWST adequately resolve the \OII doublets with an instrument resolution of $R\sim2700$. This results in the discovery of {5} galaxies in our sample, of which the \OII\ ratios are observed to be unphysical with high fidelity. An example, GLASS-20006, is presented in the lower panel of Fig. \ref{fig.40202}.

The GLASS-JWST NIRSpec data provide a unique opportunity to investigate the detectability of these unphysical \OII line ratios using the medium-resolution ($R\sim1000$) data taken by the NIRSpec/MSA medium-resolution gratings, e.g., G140M/F100LP, G235M/F170LP, and G395M/F290LP.
This test is timely, since the vast majority of the JWST NIRSpec/MSA observations were taken with the medium resolution gratings, e.g., JADES\footnote{\url{https://jades-survey.github.io/}} \citep{eisensteinJADESOriginsField2023a}, CEERS\footnote{\url{https://ceers.github.io/}} \citep{2024_Backhaus_ceersemissionlineratio}, CECILIA \citep{stromCECILIAFaintEmission2023a}, and AURORA \citep{shapley2024aurorasurveynewera}. 

\begin{figure}
    \centering
    \includegraphics[width=1\linewidth]{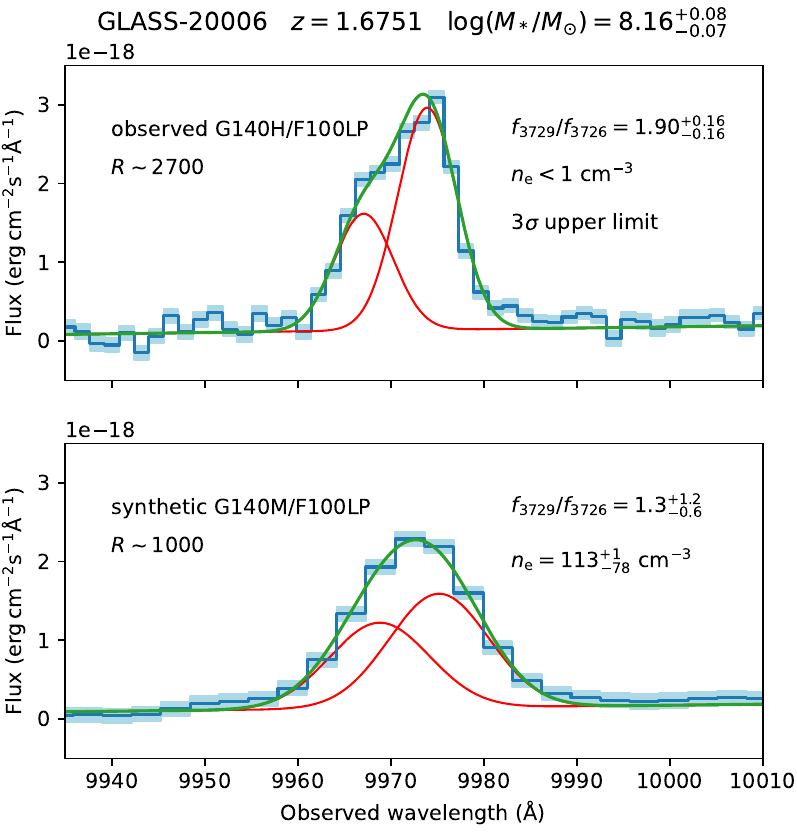}
    \caption{\label{fig.downgrading}
    Observed high-resolution ($R\sim2700$) and synthetic medium-resolution ($R\sim1000$) spectroscopy of the \OII doublets of GLASS-20006.
    The top panel shows the original spectrum of the \OII doublets secured with the high-resolution G140H grating.
    In the bottom panel, we show smoothed spectrum with effective resolution downgraded to that similar to the medium-resolution G140M grating, to which we also perform emission line profile fitting.
    We fit both doublets shown in the green Gaussian profile with their addition in green.
    We found that the value of \ne estimated from the synthetic medium-resolution data significantly deviates from that given by the original high-resolution spectrum.
    This test highlights the caveat of estimating \ne from the \OII doublets marginally resolved by the JWST NIRSpec medium-resolution spectroscopy.
    }
\end{figure}

Figure~\ref{fig.downgrading} shows an example of the impact of using medium-resolution spectroscopy on GLASS-20006. We convolved our high-resolution G140H/F100LP spectrum with a Gaussian kernel to match the $R=1000$ spectral resolution of the G140M/F100LP grating, and resampled the convolved spectra with the typical G140M/F100LP sampling to create the synthetic spectrum. The new spectrum was then fitted using the same two-Gaussian model as the high-resolution spectrum with a fixed separation and identical line widths for both components.

The synthetic medium-resolution spectrum presents a significantly different \ne. The $f_{3729}/f_{3726}$ ratio shifts from $1.90^{+0.16}_{-0.16}$  (pre-convolution) to $1.3^{+1.2}_{-0.6}$ (post-convolution), causing a change in the derived \ne(\OII) from a $3\sigma$ upper limit of $\ne <1~\mathrm{cm}^{-3}$ to $\ne = 113_{-78}^{+1}~\mathrm{cm}^{-3}$. With medium-resolution spectroscopy, this galaxy would have been misclassified as an ordinary galaxy with a reasonable \ne\ measurement.

\subsection{SED Fitting}   \label{subsec:m_sfr}

We performed SED fitting to derive \mstar for each galaxy.  The photometric galaxy fluxes used in the SED fitting were acquired from the photometric catalog constructed by DJA.
We adopt the fluxes observed by both JWST and HST.

Our fitting methodology follows that of \cite{Santini_2023}, employing the Bayesian SED-fitting code \textsc{\tt Bagpipes} \citep{2018Carnall} and the 2016 version of the stellar population models by \cite{Bruzual2003}.  The nebular continuum and emission lines were modeled using the photoionization code \textsc{\tt CLOUDY} \citep{2017FerlandRMxAA..53..385F}, following the methodology developed by \cite{2017Byler}. We assumed a \cite{Chabrier2003} initial mass function (IMF) and double-power-law star formation histories, applying the \cite{Calzetti2000} dust attenuation law.

The GLASS-JWST field is gravitationally lensed by the foreground cluster A2744. Consequently, it is necessary to account for the lensing magnification in order to accurately determine the physical parameters of each galaxy. To achieve this, we use the magnification model \textsc{CATSV4.1} developed by \cite{2015Jauzac}, which allows us to calculate the magnification of each source based on its location. This magnification is then corrected for when deriving \mstar, as well as for the SFR (Section~\ref{subsec:balmer_sfr}).

We compare our derived \mstar values with those obtained by \citet{2024_Merlin} using photometric $z$ and found good agreement, with a median difference of {0.03\,dex} and scatter of 0.1\,dex. The scatter is primarily due to the variance between their photometric redshifts and our more robust spectroscopic redshifts. We caution that at $z\gtrsim 4.7$, the F444W band fails to observe rest-frame NIR wavelengths where fluxes are dominated by old stars, potentially making stellar mass derivations for these high-$z$ galaxies less reliable. Our \mstar\ measurements are summarized in Table~\ref{tab:ne}. We refrain from using SFRs from the SED fitting due to their unreliability caused by the lack of rest-frame UV imagings in some of our galaxies. Instead, we measure SFRs using Balmer lines, as discussed in the next section.

\subsection{Balmer-line Based SFR}   \label{subsec:balmer_sfr}

\begin{figure}
    \includegraphics[width=1\linewidth]{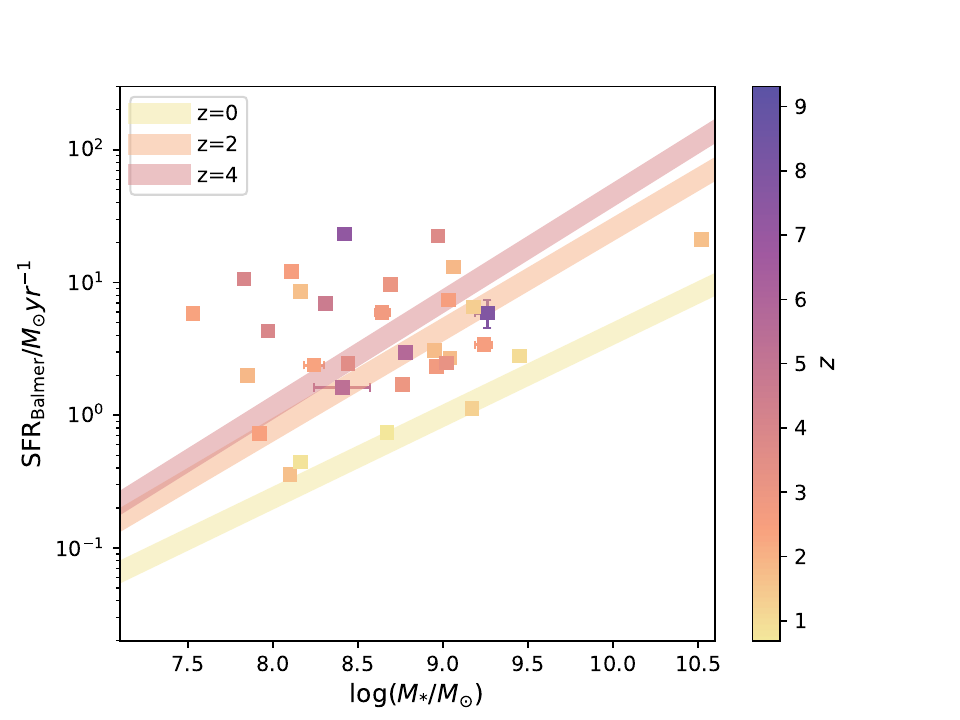}
    \caption{\label{fig.sfr_m}
   The distribution of \mstar\ and SFR for the galaxies in our sample. The data points are color-coded by their redshifts. The solid thick lines indicate the star-forming main sequence at the corresponding redshifts adopted from \citet{Speagle2014}. Our galaxy samples are represented by large squares, and are roughly representative of the star-forming main sequence at the corresponding redshifts.
    }
\end{figure}

We estimate the SFR for each galaxy using the Balmer emission lines, following the calibration proposed by \cite{kennicutt98} with the \cite{Chabrier2003} IMF:
\begin{align}\label{eq.sfr}
   {\rm SFR = 1.3 \times  10^{-41}} \frac{L(\Hb)}{\rm [erg \, s^{-1}]} [M_{\odot}\,{\rm yr^{-1}}] ,
\end{align}
where $L(\Hb)$ represents the intrinsic luminosity of the \Hb line after correcting for dust extinction and lensing magnification. In cases where \Hb is not available due to wavelength coverage limitations, we estimate the \Hb flux by converting the \Ha flux using the theoretical ratio of $L(\Ha)/L(\Hb) = 2.86$ in the case B recombination scenario. To correct for dust extinction using the Balmer decrement, at least two significantly detected Balmer lines are required. However, in our sample, 5 out of \Ntot galaxies does not have reliable Balmer decrement measurements. To maximize the number of galaxies in our analysis, we employ the SED-derived $A_{V}$, which is available for all galaxies in our sample. We verified that the SED-derived $A_V$ values are consistent with those derived from the Balmer decrement. 

\begin{figure}
    \centering
    \includegraphics[width=0.95\linewidth]{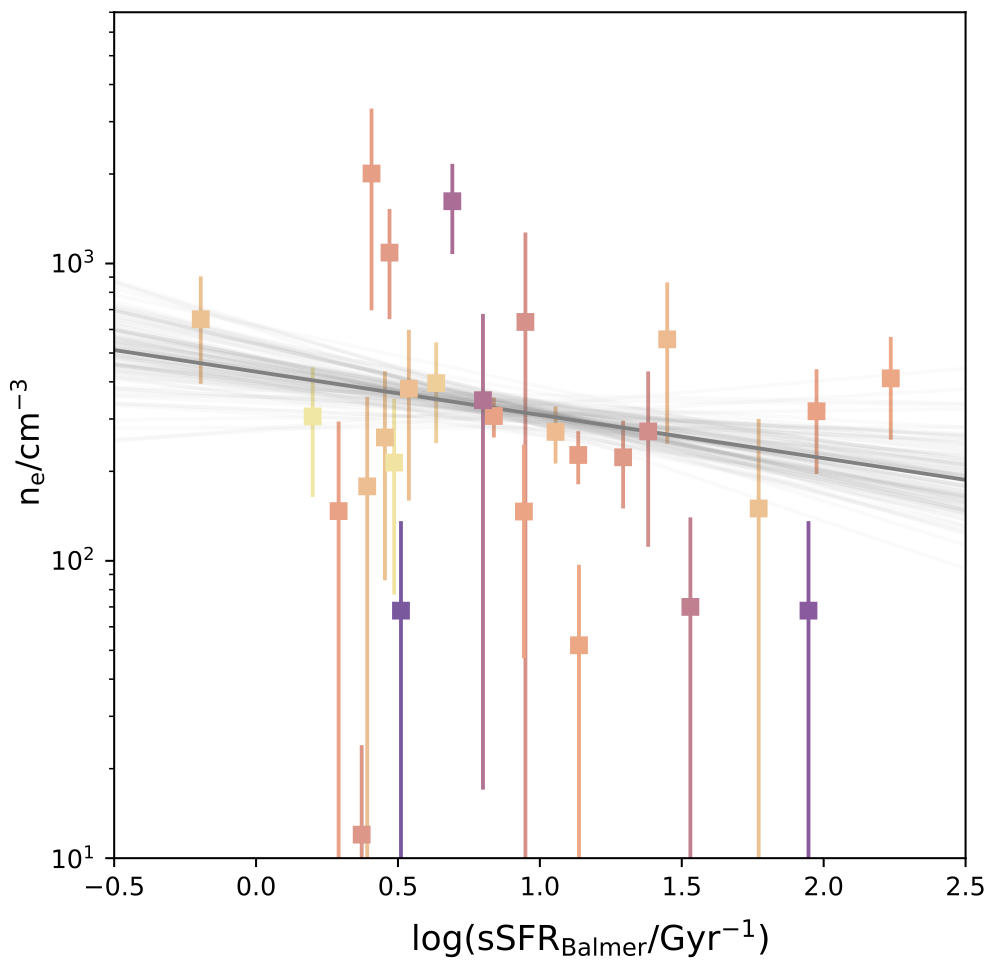}  
    \caption{\label{fig.ne_sfr}
    The relationship between \ne and sSFR of all galaxies in our sample. The color coding represents $z$, with the same scale shown in Fig.~\ref{fig.sfr_m}. The thin gray lines mark 50 random draws from the linear regression, with the dark gray line representing the best-fit.
    }
\end{figure}

 \begin{figure*}[htbp]
    \centering
    \includegraphics[width=0.95\linewidth]{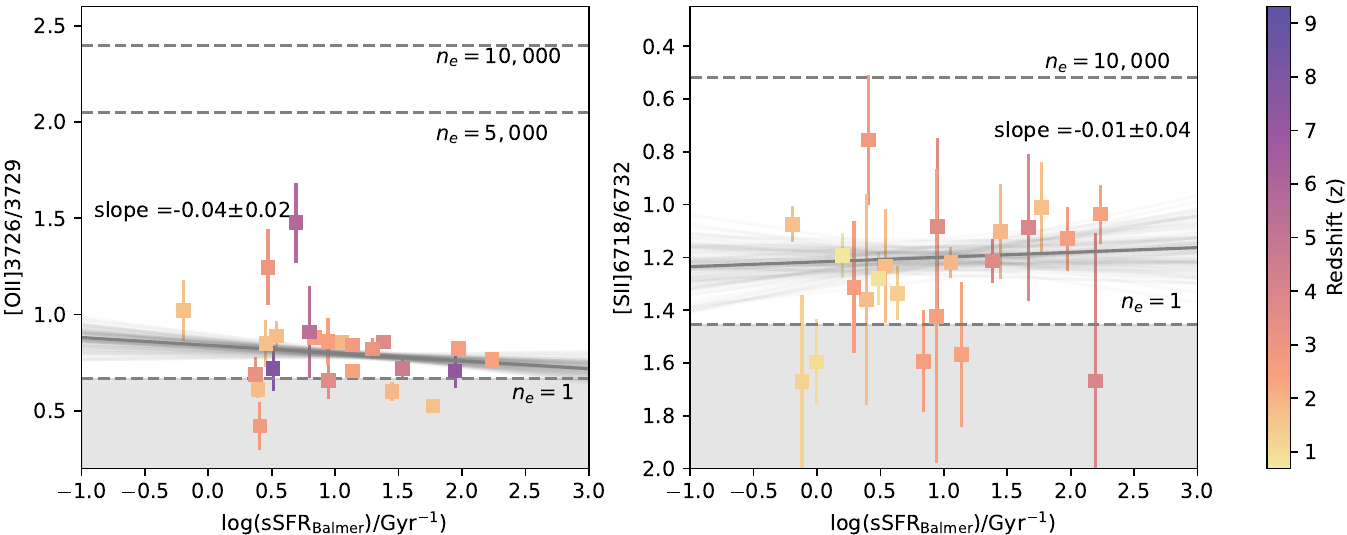}
    \caption{\label{fig.r_logssfr}
    The relationships between line flux ratios and specific star formation rate (sSFR) of our sample galaxies.
    The left panel shows the \OII$\lambda3726/ \OII \lambda 3729$ flux ratio, and the right panel shows the $\SII\lambda 6718/\SII \lambda 6732$.
    Note that the \OII$\lambda\lambda3726,3729$ doublets are well resolved by our GLASS-JWST NIRSpec MOS spectroscopy taken with the high-resolution gratings.
    The SFR values shown in both panels are measured from the reddening corrected Balmer line fluxes. The color coding represents $z$ as shown by the color bar on the right. 
    In both panels, the dark and thin gray lines corresponding to the best-fit and 50 random draws from the linear regression. 
    The p-value in the left and right panel is 0.22 and 0.44, respectively.
    }
\end{figure*}
The extinction $A_V$ is used to derive $E(B-V)_\mathrm{SED}$, using the relation $R_{V} \equiv \frac{A_{V}}{E(B-V)}$ with $R_{V} = 4.05$ following the \cite{Calzetti2000} dust law. We then apply the correction $E(B-V)_\mathrm{Balmer} = (0.44 \pm 0.03) \times E(B-V)_\mathrm{SED}$, as suggested by \citet{2019Theios_KBSS} for the model combing 
\citet{Bruzual2003} and \citet{Calzetti2000} SED model.
The resulting $E(B-V)_\mathrm{Balmer}$ is then used to correct for the dust attenuation in the SFR calculation.  Our derived $A_V$ and SFR are listed in Table~\ref{tab:ne}. We compared SFR and \mstar of our samples with the star-forming main sequence (SFMS) derived by \cite{Speagle2014} in Figure~\ref{fig.sfr_m}. We found that our samples are consistent with star-forming galaxies, showing higher SFR for galaxies at higher redshift.

\section{Results and Discussions}     \label{sec:discu}

\subsection{\ne vs. sSFR}  \label{subsec:ne_sSFR}
Previous studies have indicated a connection between \ne\ and star formation {activity}. \citet{reddyJWSTNIRSpecExploration2023} demonstrated that the star formation surface density ($\Sigma_{\rm SFR}$) is correlated with the \SII $\lambda\lambda 6718/6732$ flux ratio in galaxies at $z=2.7$--6.3. This relation may be the result of the Kennicutt-Schmidt relation \citep{kennicutt98}, where a dense ISM plays a crucial role in promoting star formation. This is further supported by the observed correlation between the ionization parameter and $\Sigma_{\rm SFR}$ \citep{2023Reddy_Sigmasfr&U_MOSFIRE}. 
However, measuring galaxy sizes at high redshift is challenging, as these galaxies often exhibit irregular structures that significantly deviate from the assumed Sersic profile used in 2D fitting, resulting in large uncertainties. Furthermore, the wide redshift range of our sample implies that JWST NIRCam does not consistently capture a rest-frame size, which is crucial for accurately characterizing galaxy size \citep{van_der_Wel_2014}. Instead, we probe the dependence on sSFR. The UV radiation from a large number of massive stars in intensely star-forming galaxies may ionize more atoms in the gas, releasing more free electrons and thereby increasing the electron density. sSFR is preferred over SFR because we aim to use mass as a proxy for size, recognizing that more massive galaxies typically have larger sizes \citep{van_der_Wel_2014}, while also attempting to remove any mass dependence.

\begin{figure}[htbp]
    \includegraphics[width=1\linewidth]{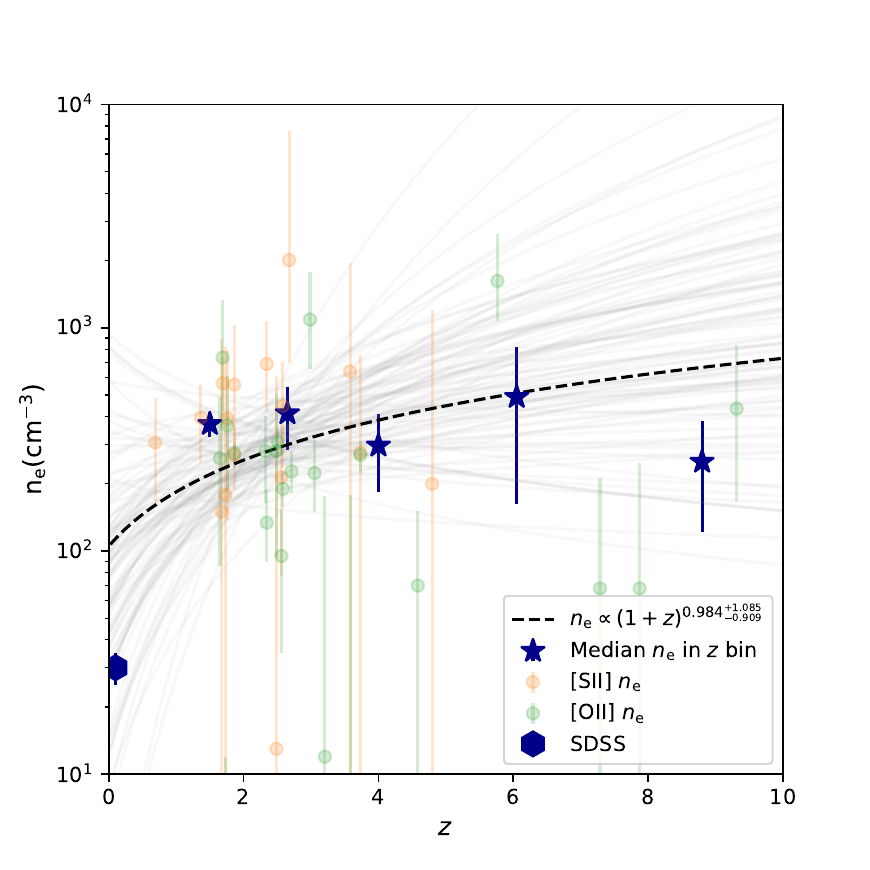}
    \caption{\label{fig.ne_z}
    Redshift evolution of the ISM electron density (\ne) of the high-redshift galaxies analyzed in this work.
    The \ne measurements based on the low-ionization emission line flux ratios of the \OII and \SII doublets are represented by the green and orange circles, respectively.
    The blue stars correspond to the running median \ne values measured in the five redshift bins. The hexagon shows the median measurement from the SDSS galaxies at $z\sim0$.
    The black dashed line marks our best-fit redshift evolution with 1-$\sigma$ uncertainties of 
    $\ne\propto(1+z)^{0.984^{+1.085}_{-0.909}}$
    to our measurements at high redshifts and the SDSS measurement at $z\sim0$.
    The thin gray lines denote 50 random draws based on the 1-$\sigma$ uncertainties of the linear regression.
    We note that our constrained relation between \ne and $z$ is also compatible with no redshift evolution of \ne across $z\sim2$ to $z\sim9$.
    }
\end{figure}
Figure~\ref{fig.ne_sfr} compares the \ne\ with sSFR for the galaxies in our sample. However, we do not find a significant linear correlation between \ne and sSFR.  
{By doing linear regression, the best-fit slope yields $-0.15^{+0.07 }_{-0.06 }$. The $p$ value of Pearson correlation coefficient is 0.47, indicating that the linear correlation between sSFR and \ne is not statistically significant. 
}

To further investigate the lack of correlation between \ne\ and sSFR, we split the sample into the ones with \OII\ measurements (the ``\OII'' + ``overlap'' samples) and the ones with \SII\ measurements (the ``\SII'' + ``overlap'' samples), and compared the relationship between the doublet ratios and sSFR (Fig. \ref{fig.r_logssfr}). By doing linear regression between \OII ratio and $\rm log(sSFR/Gyr^{-1})$, the best-fit slope yields $-0.04\pm {0.02 }$ with $p$ value = 0.22. For \SII the best-fit slope is $-0.01 \pm 0.04$ with $p$ value = 0.44.

Although these high $p$ values suggest that the two correlations are not statistically significant,
the marginal correlation between the \SII\ ratio and sSFR \tcb{may} echo the \SII\ ratio vs. $\Sigma_{\rm SFR}$ correlation discovered in \citet{reddyJWSTNIRSpecExploration2023}. In contrast, the reversion correlation between the \OII\ ratio and sSFR may be surprising since the errors assocaited with the \OII\ flux ratios are typically $\sim 3$ times lower than the errors of the \SII\ ratios. 
To test whether this discrepancy is caused by the redshift differences between the two samples, we removed galaxies with $z > 6$ in the \OII\ sample, and still found a similar correlation between the \OII\ ratio and sSFR. Therefore, our results \tcb{may} suggest that the \ne\ derived from \OII\ does not reflect the similar correlation found for \SII. In \S\ref{subsec:oii_sii}, we will continue the discussion of the comparison between the \OII\ and \SII\ \ne.

\subsection{\ne vs. $z$}  \label{subsec:ne_z}
The potential redshift evolution of \ne has been discussed by \citet{Davies2021,isobe2023redshift,2024Abdurro'uf_nez} up to $z \sim 9$. Using medium resolution spectra of the \OII\ doublet, they found that $n_{\rm e} = (1+z)^k$, where $k\simeq 1$--2. This connection could be the result of the redshift evolution of galaxy size \citep[e.g.,][]{van_der_Wel_2014}, so that higher-$z$ galaxies have denser ISM simply because they are more compact.

Figure~\ref{fig.ne_z} plots the dependence of our measured \ne on $z$. To investigate this redshift evolution in our sample, we divide our samples into five bins, and calculate the median \ne\ in each bin, marked by blue stars. Based on these median values, we derive the power-law index by fitting them to the $n_{\rm e} = (1+z)^k$ function using maximum likelihood estimation (MLE) through the MCMC method using \textsc{emcee} \citep{Foreman_Mackey_2013}.  Through our sample, we did not find significant connection between \ne\ and $z$ for our sample, with the best-fit $k=0.48^{+0.44}_{-0.42}$.

We further include an average \ne\ at $z=0$, marked by blue hexagon in Fig.\ref{fig.ne_z}, measured by \cite{Kaasinen2017}. \cite{Kaasinen2017} measured the \ne based on the measurement of emission lines of spectra from the Sloan Digital Sky Survey (SDSS) provided by \cite{Kauffmann2003}. The median stellar mass of their sample follow $\log(M_*/M_{\odot})=10.6$. Note that the insufficient number of high-$z$ samples makes it difficult to ensure sample similarity. By including the $z\approx0$ measurement, we re-fit the correlation between \ne and $z$, and plot the best-fit function as black dashed curve in Fig.~\ref{fig.ne_z}. This increases the power-law index to $k = 0.98 ^{+1.09}_{-0.91}$.

While the measurements of \ne\ show a clear jump from $z\simeq0$ to 2, it is not enough to offset the lack of correlation at $z\simeq 2$--10. Compared to \citet{isobe2023redshift}, the \ne results measured in our sample at $z\simeq 2$--3 are about twice their value. This might be caused by the difference in galaxy mass. 
\tcb{However, we tested the dependence of \ne on stellar mass and did not find a correlation for them. }
Although \citet{isobe2023redshift} adopted the literature \ne measured from \SII or \OII, similar to our sample, without JWST and lensing magnification, these galaxies have a typical $M_* \sim 10^{10}~M_\odot$ that is larger than all our galaxies. 
For galaxies at $z\simeq 6$--10, our measurements are only $\sim 0.2$--0.3 in \ne\ compared to \citet{isobe2023redshift}. Since there are only 3--4 galaxies in this redshift range in both works, this could be caused by the small-sample variation. In conclusion, using the internally consistent sample with similar $M_*$ across $z = 1.5$--10, we detect no significant redshift evolution of \ne. 
\tcb{This may arise from the declining density of H\,\textsc{ii} along with the declining SFR, which peaks at cosmic noon \citep{Madau2014}}

\begin{figure*}
    \centering
    \includegraphics[width=1\linewidth]{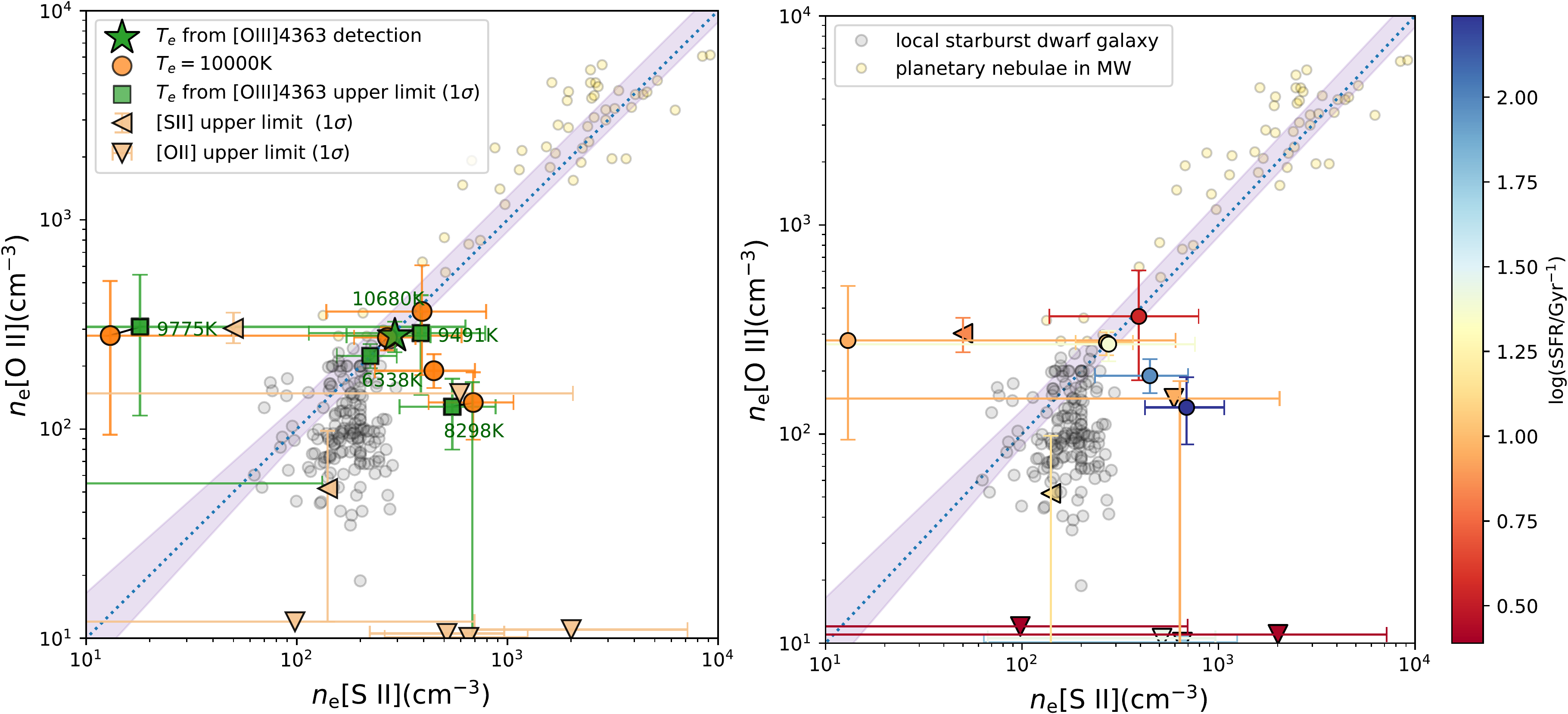}
    \caption{
    Comparison of the ISM electron density derived from the \OII$\lambda\lambda$3726,3729 and \SII$\lambda\lambda$6718,6732 doublet flux ratios.     In the left panel, the orange circles denote \ne measured using both tracers, assuming a fiducial electron temperature of \te(\OII)=\te(\SII)=10,000 K, with the error bars showing 1-$\sigma$ uncertainties. The 1-$\sigma$ upper limits are marked by triangles. 
    We also estimate \te(\OII) from the detection or 1-$\sigma$ upper limit of the \OIII$\lambda$4363 auroral line using Eq.~\ref{eq.te}, listed in green numbers, and then derive \ne constraints, marked in green symbols.    In both panels, the gray points show the comparison for planetary nebulae in the Milky Way \citep{1992Kingsburgh}, the yellow dots mark the \ne comparison in individual \HII regions within a local starburst galaxy \citep{2023Fernandez}, and the purple shaded band corresponds to the 1-$\sigma$ confidence interval of the \ne comparison of a sample of local \HII regions analyzed in \citet{Sanders2016}.     In the right panel, we instead color code our results using their corresponding sSFR measurements (see Sect.~\ref{subsec:balmer_sfr}).
    }
    \label{fig.ne_Te}
\end{figure*}

\subsection{Comparing \ne\ from \OII\ and \SII}   \label{subsec:oii_sii}
In Sect.~\ref{subsec:ne_sSFR}, we have shown that using \OII\ and \SII\ ratios could yield different correlations in \ne. 
\tcb{
Different diagnostics have been found to give different densities. 
In particular, \OII\ is more sensitive to slightly higher densities \citep{2022Mingozzi}, perhaps causing inconsistent \ne calculated based on \OII\ and \SII. Our data are superior for resolving the \OII\ doublet among all currently existing NIRSpec/MSA spectroscopy.}
In this section, we investigate whether the two \ne\ indicators are intrinsically consistent in our sample. 
Similar analyses has been conducted for local star-burst galaxies \citep{2023Fernandez} and for $z \sim 2$ galaxies by \citet{Sanders2016}. For $z\sim 2$ galaxies, \citet{Sanders2016} found that $n_{\rm e}$(\OII) and $n_{\rm e}$(\SII) are consistent within $\sim 20\%$. While \citet{2023Fernandez} found that $n_{\rm e}$(\SII) are higher than $n_{\rm e}$(\OII) {on average}. 

In Fig. \ref{fig.ne_Te}, we present the correlation between $n_{\rm e}$(\OII) and $n_{\rm e}$(\SII). The robust measurement with \te of 10,000 K are marked by orange dots, and those upper limit are marked by triangle. Those for local star-burst galaxies \citep{2023Fernandez} are marked by gray dots, and those for planetary nebulae in the Milky Way \citep{2006Osterbrock} are marked by light yellow dots. The dotted line and its associated purple shaded region corresponds to the best-fit function and 1-$\sigma$ confidence level from the \ne comparison of a sample of local \HII regions analyzed by \cite{Sanders2016}.  Although our sample contains only \Novl\ galaxies, it is evident that our $n_{\rm e}$(\OII) vs. $n_{\rm e}$(\SII) relation has a much larger scatter compared to the other samples. 
\tcb{This large scatter partly comes from measurement uncertainties, }
and is partly driven by large amount of galaxies with upper limits in \ne\ from the unphysical flux ratios. 
Excluding the upper limits, the large scatter persists due to one galaxy with low $n_{\rm e}$(\OII) but significantly high $n_{\rm e}$(\SII).

Although the \OII\ and \SII\ doublets are mostly sensitive to \ne, thus, making them optimal \ne\ indicators, $T_e$ could still have small but non-negligible effect on the flux ratios. To further remove this systematic uncertainty, we include one robust $T_e$ measurement and upper limits of {five} galaxies in the overlap sample and present the resulting \ne\ {as green symbols} in Fig. \ref{fig.ne_Te}. As expected, while $T_e$ changes the $n_{\rm e}$(\OII) and $n_{\rm e}$(\SII) slightly, its effect is inadequate to reconcile the large scatter.

The scatter could also be connected to the sSFR as we show in the {right panel} of Fig. \ref{fig.ne_Te}. {3/\Novl} galaxies with $\log(\rm sSFR/Gyr^{-1})\gtrsim$1 and 1 galaxy $\log(\rm sSFR/Gyr^{-1})=$ 0.85 are inconsistent to the one-to-one ratio by 1-$\sigma$, while { $\log(\rm sSFR/Gyr^{-1})\lesssim$0.8} has all \ne consistent within 1 $\sigma$. 
This echoes the consistently large $n_{\rm e}$(\SII) measured in local star-burst galaxies, where feedback from rapid star formation could drive larger density perturbations in the ISM. Since S$^+$ has a lower ionization energy (23.33 eV) compared to O$^+$ (35.12 eV), \SII\ emission could be biased toward low-$T_e$ and high \ne\ gas. These effects are amplified in the high-$z$ universe, where galaxies experience stronger and often bursty star formation \citep[e.g.][]{2018Faucher-Giguere_starburst, 2023sun_starburst_jwst}, causing the $T_e$ and \ne\ structures of \HII\ regions in these galaxies more complex. Our results fit into the complex nature of \HII\ regions supported by temperature fluctuations \citep{2002Peimbert_te_in_Gaseous_Nebulae, 2019Peimbert_te, 1993Peimbert} and the recent observational evidence for density fluctuations \citep{md23}. Our result also suggests that processes driven by star-formation (e.g., stellar feedback, turbulence, and shock heating) may play a significant role in creating density fluctuations.
More spectroscopic observations with similar setup to that of the NIRSpec component of GLASS-JWST are needed to increase the sample size of such detailed comparisons between $n_{\rm e}$ measured from both doublets, in order to provide a more definitive characterization of the high-$z$ ISM structure at high statistical significance.

\section{Conclusions}   \label{sec:conclu}

In this work, we provided \ne\ measurements in a sample of \Ntot galaxies using the high-resolution NIRSpec/MSA data acquired by the GLASS-JWST project. These galaxies span a wide redshift range of $0.7<z<9.3$ and have stellar masses of $\log (M_* / M_\odot) \simeq 7.5$--10.5 across all redshifts, thanks to the sensitivity boost from lensing magnification introduced by the foreground A2744 cluster. The \ne\ were measured from the \OII\ $\lambda\lambda$3727/3729 and \SII\ $\lambda\lambda$6716/6731 ratios, and compared to the redshift and sSFR of the host galaxies. Our results are summarized below.

\begin{itemize}

   \item There is positive correlation between the $\OII \lambda\lambda3726/3729$ ratio and sSFR and a negative correlation between the $\SII \lambda\lambda6718/6732$ ratio and sSFR, supporting the previously discovered connection between \ne\ and star formation. However, the \OII ratio and the \SII ratio has opposite trends as the sSFR increases. That means that the $\ne[\OII]$ decreases with the sSFR increases while the $\ne[\SII]$ increases as the sSFR increases.
   Meanwhile, the combined \ne\ measured from \OII\ and \SII\ show a negative correlation between \ne\ and sSFR, indicating that different gas may have varying sensitivity and activity in response to the star-formation activities in \HII\ regions. 
   
   \item From redshift $z\approx0$ to $z\approx2$, \ne significantly increases from $\sim 30~\mathrm{cm}^{-3}$ to $\sim 400~\mathrm{cm}^{-3}$. 
   However, there is no obvious redshift evolution of \ne\ in the range of $z \approx 2$--10 in our sample.  

   \item In the sample of \Novl galaxies where both \ne(\OII) and \ne(\SII) can be measured, there is a much larger scattering in \ne(\OII) vs. \ne(\SII) compared to the local samples of star-burst galaxies and planetary nebulae. Galaxies that deviate more significantly from the one-to-one ratio may be associated with larger sSFR, supporting the idea that star formation feedback drives complexity in \HII\ regions.   

   \item From the high-resolution ($R \sim 2700$) spectroscopy from JWST/NIRSpec, we discovered 4 galaxies with a \OII\ line ratio that exceeds the theoretical upper limit by $>3\sigma$. By simulating observations of these galaxies from the medium-resolution ($R \sim 1000$) JWST/NIRSpec observations, we found that {the measured line ratio is significantly underestimated}, and thus the intrinsic unphysical line ratio cannot be recovered.  This highlights the importance of high spectral resolution in \ne\ measurements. 

\end{itemize}

Our study provides valuable insight into the properties of high-$z$ emission line galaxies and their relationship with redshift and sSFR. Our sample also underscores the importance of using similar observations in high-sensitivity and high spectral resolution to study the ISM environment in high-$z$ galaxies, particularly by JWST. However, the sample size of our sample is limited to only $\sim \Ntot$ galaxies, increasing the sample size of which \ne\ can be directly measured in high-$z$ galaxies with future observations will be important to confirm our results.

\begin{acknowledgements}
We thank the anonymous referee for valuable suggestions. We thank Si-Yue Yu, Yuki Isobe, and {David {Fern{\'a}ndez-Arenas}} for useful discussions.
This paper is dedicated to the memory of our beloved colleague Mario Nonino who passed away prematurely. We miss him and are indebted to him for his countless contributions to the GLASS-JWST project.
XW is supported by the National Natural Science Foundation of China (grant 12373009), the CAS Project for Young Scientists in Basic Research Grant No. YSBR-062, the Fundamental Research Funds for the Central Universities, the Xiaomi Young Talents Program, and the science research grant from the China Manned Space Project.
XW also acknowledges the work carried out, in part, at the Swinburne University of Technology, sponsored by the ACAMAR visiting fellowship.
HRY is supported by National Science Foundation of China grant No. 12173030.
This work is based on observations made with the NASA/ESA/CSA JWST, associated with program JWST-ERS-1324.
{The JWST data presented in this article were obtained from the Mikulski Archive for Space Telescopes (MAST) at the Space Telescope Science Institute. The specific observations analyzed can be accessed via \dataset[DOI:10.17909/kw3c-n857]{https://doi.org/10.17909/kw3c-n857}.}
The MAST at the Space Telescope Science Institute, which is operated by the Association of Universities for Research in Astronomy, Inc., under NASA contract NAS 5-03127 for JWST. 
We acknowledge support from the INAF Large Grant 2022 “Extragalactic Surveys with JWST”  (PI Pentericci). K.G. and T.N. acknowledge support from Australian Research Council Laureate Fellowship FL180100060. BV is supported  by the European Union – NextGenerationEU RFF M4C2 1.1 PRIN 2022 project 2022ZSL4BL INSIGHT. 
\end{acknowledgements}

\bibliographystyle{aasjournal}

\end{document}